\documentclass[12pt]{article}
\def\rt2{\frac{1}{\sqrt{2}}} 
\begin{document} 
%\hfill{preprint No} 
\begin{center} 
{\LARGE \bf Comments related to reading "{\it Static quantities of the $W$
boson in the $SU_L(3)\times U_X(1)$ model with
right-handed neutrinos} "}\\[1cm] {\bf Nguyen Anh Ky}\\[4mm] 
Institute of Physics, P.O. Box 429, Bo Ho, Hanoi
10 000, Vietnam
%\\[4mm] and\\[4mm]
%Abdus Salam International Centre for Theoretical Physics, 
%34 100 Trieste, Italy  
\end{center} 
\vspace{2mm} 
\begin{abstract} 
Comments related to reading the paper "{\it Static quantities of the $W$ 
boson in the $SU_L(3)\times U_X(1)$ model with right-handed neutrinos}", 
hep-ph/0312308 (by J. L. Garcia-Luna et al) are given.  They do not concern 
the main results of the paper but a statement there about the Higgs sector of the 
331 model with right-handed neutrinos. The scalar sextet introduced here may 
help us to generate neutrino masses in an acceptable range. 
\end{abstract} 

It is stated in  \cite{garcia} that in comparison with the minimal version of the 
331 model the version with right-handed neutrinos (or shortly, the RHN model) 
requires a simpler Higgs sector, more precisely,  the latter consists of three Higgs 
triplets only and no a sextet is needed (in order to reproduce the known physics at the 
Fermi scale). However, such a simple Higgs sector may not be enough when we 
want to make the neutrinos massive. The RHN model, in the present status, cannot 
explain (at least, at the tree-level) the smallness of neutrino masses and it does not 
generate a Majorana neutrino mass which, according to our opinion, should not be 
exluded in advance from  consideration. This problem  may be solved by introducing 
a Higgs sextet to the model.\\  

  The minimal 331 model can generate Majorana masses which could be 
tree-level (if the neutral sextet component $S_{00}\equiv \sigma_1^0$ 
develops a non-zero vacuum expectation value)  \cite{tj} or radiatively 
induced \cite{kitabay,okamoto}. There are also other mechanisms for 
generating neutrino masses in this version of the 331 model such as the 
one of \cite{mpp2} where the sextet is replaced by a neutral scalar singlet 
and a dimension-seven effective operator is considered. Experimental and 
practical data show that the neutrinos if massive have very tine masses 
(only a few eV's or less). In the frameworks of the  minimal 331 model
some attempts for explaining the smallness of neutrino masses have been made
(see, for eaxmple, \cite{tj} -- \cite{mpp1}, \cite{mpp2}). As far as the 331 model 
with RHN's is concerned, investigations on neutrinos masses here are much poorer.
In this model, the neutrinos (more precisely, two of them) can gain Dirac masses 
if a definite one of the three Higgs triplets develops a non-zero VEV but at the 
present, as stated in \cite{flt}, it is not known how to get small neutrino masses. 
Additionally, there is no reason the Majorana neutrino masses to be excluded 
from consideration when the lepton number has no real meaning in both versions 
of the 331 model (as a lepton and its antiparticle are simultaneosly components of 
one and the same multiplet) and when neutrinoless double beta decays (which are 
sensitive to the  existence of Majorana neutrinos and violate the total lepton number 
by two uinits) are still considered as possible processes. Moreover, the presence of 
independent right-handed neutrinos in the theory is a good reason for considering 
neutrino masses of both Dirac and Majorana types. \\

The Higgs sector of the original version with RHN's \cite{flt} consists of three scalar 
triplets 
\begin{equation}
\eta= \left( \begin{array}{c}
\eta^0 \\[2mm] \eta_1^- \\[2mm] \eta_2^+ 
\end{array} \right) 
\sim (1,3,-1/3),~~ 
\rho= \left( \begin{array}{c}
\rho^+ \\[2mm] \rho^0 \\[2mm] \rho^{++}
\end{array} \right)
\sim(1,3,2/3),~~
\chi= \left( \begin{array}{c}
\chi^- \\[2mm] \chi^{--} \\[2mm] \chi^0 
\end{array} \right)
\sim (1,3,-1/3).
\end{equation}
The Yukawa couplings in this case 
\begin{eqnarray}
%\[
{\cal L}^{\chi}_{Yuk} &=& \lambda_1 \bar Q_{1L} u^{'}_{1R} \chi
+ \lambda_{2ij} \bar Q_{iL} d^{'}_{jR} \chi^{\dagger} + H.c., 
%\]
\label{ykw}
\nonumber \\[2mm]
%\[ 
{\cal L}_{Yuk}^{\rho} &=& \lambda_{1a} \bar Q_{1L} d_{aR} \rho
+ \lambda_{2ia} \bar Q_{iL} u_{aR} \rho^{\dagger} + G_{ab}
\bar f^a_L(f^b_L)^c \rho^{\dagger} + G^{'}_{ab} \bar f_L^a e_R^b \rho + H.c., 
%\]
\\[2mm]
%\[
{\cal L}^{\eta}_{Yuk} &=& \lambda_{3a} \bar Q_{1L} u_{aR} \eta +
\lambda_{4ia} \bar Q_{iL} d_{aR} \eta^{\dagger} + H.c.,
%\]
\nonumber
\end{eqnarray} 
where $a,b=1,2,3$, $i=2,3$, can ensure 
masses  for all quarks and charged leptons as well as Dirac masses for two of  
the neutrinos \cite{flt}. This model, at least, at the tree-level, however, cannot 
explain the smallness of neutrino masses and it does not generate  a Majorana 
neutrino mass which should be by no reason exluded in advance from 
consideration. \\

A neutrino mass (at the tree level) can be generated by coupling an appropariate 
Higgs boson to $\bar f_L(f_L)^c$ with $f_L$ being a lepton multiplet of given 
generation having the following components and tranformation rule under the 
331-gauge group $SU(3)_c\times SU(3)_L\times U(1)_X$   
\begin{equation}f^a_L =
\left(\begin{array}{c} \nu_L^a\\[2mm] e_L^a\\[2mm]
(N_R^c)^a
\end{array}\right) \sim (1, 3, -1/3), \  e_R^a \sim (1, 1, -1),
\end{equation}
where $a = 1,2,3,$ is a family index, $N_R$ is a right-handed neutrino different 
from the anti-neutrino $\nu_R$ -- the anti-particle of $\nu_L$ (to avoid any confusion 
we use here another notation for the third component of $f_L$ instead the one used 
in \cite{flt}). As an $SU(3)_L$-representation state $\bar f_L(f_L)^c$ is a tensor 
product ${\bf 3}^{*}\otimes  {\bf 3}^{*}$ of two $SU(3)_L$--anti-triplet $\bar f_L$ 
and $(f_L)^c$, consequently, it can be decomposed into a direct sum of a triplet {\bf 3} 
(the anti-symmetric part of the tensor) and an anti-sextet {\bf 6}$^{*}$  (the symmetric 
part of the tensor): $${\bf 3}^{*}\otimes {\bf 3}^{*}= {\bf 3}\oplus {\bf 6}^{*}.$$
To constitute an $SU(3)_L$-invariant quantity we could contract  
$\bar f_L(f_L)^c$ with an anti-triplet ${\bf 3}^{*}$ and/or a sextet 
{\bf 6}. The term $G_{ab}\bar f^a_L(f^b_L)^c\rho^{\dagger}$ in 
the Yukawa Lagrangians (\ref{ykw}) is a contraction of the first kind. 
This Yukawa coupling term, at $\langle \rho \rangle \neq 0$, can generate Dirac 
masses for two of the three neutrinos (while the third  one remains massless) \cite{flt}. 
This way of  generating neutrino masses, as mentioned above, gives no indication for that the 
neutrino masses obtained are small, and it excludes the Majorana neutrino masses 
which might be important. Adding a scalar sextet to the Higgs sector may help us 
to solve these problems.\\

   A scalar filed tranforming under $SU(3)_L$ as a sextet {\bf 6} can be 
described by a symmetric tensor which in the present case has the following
explicit form and 331-gauge transformation 
\begin{equation}
{\cal S}= \left( \begin{array}{ccc}
\tau_1^0 & ~ T_1^-/\sqrt{2} & ~ \tau_2^0 /\sqrt{2}\\[3mm]
T_1^- /\sqrt{2} & ~ T_2^{--} & ~ T_3^- \\[3mm]
\tau_2^0 /\sqrt{2}& ~ T_3^- & ~ \tau_3^0
\end{array} \right)
\sim (1,6,-2/3).
\end{equation}
\\
A non-zero VEV of this sextet ${\cal S}$ coupled to the symmetric part of 
$\bar f_L(f_L)^c$ could give rice to Dirac and/or Majorana neutrino masses
without changing (effecting) the masses of the charged leptons. A Lagrangian
term corresponding to this Yukawa coupling is   
\begin{equation}
G^s_{ab}\bar f^a_L(f^b_L)^c{\cal S} + \mbox {H. c.}, 
\label{ykws}
\end{equation}
where $G^s_{ab}$ are new coupling constants with $a,b=1,2,3$, being family 
indeces, while the $SU(3)_L$-indeces are implicit. \\ 

A general structure
of a VEV of ${\cal S}$ could be  
\begin{equation}
\langle {\cal S}\rangle = \left( \begin{array}{ccc}
\omega_1 & ~ 0  & ~ \omega_2 /\sqrt{2}\\[3mm]
0 & ~ 0 & ~ 0 \\[3mm]
\omega_2 /\sqrt{2}& ~ 0 & ~ \omega_3
\end{array} \right), 
\end{equation}
\\
where $\omega_i$ are VEV's of the neutral sextet components $\tau^0_i$,
$i=1,2,3$. This VEV when inserted in (\ref{ykws}) leads to the mass
term \\
\begin{equation}
G^s_{ab}\left(\bar {\nu}_L^{~a}, ~ \bar e_L^{~a}, ~ 
(\bar N_R^{~c})^a\right)
\left( \begin{array}{ccc}
\omega_1 & ~ 0  & ~ \omega_2 /\sqrt{2}\\[3mm]
0 & ~ 0 & ~ 0 \\[3mm]
\omega_2 /\sqrt{2}& ~ 0 & ~ \omega_3
\end{array} \right)
\left(\begin{array}{c} (\nu_L^{~c})^b\\[2mm] (e^c_L)^b\\[2mm]
N_R^{~b}
\end{array}\right),  
\end{equation}
\\
which in the neutrino subspace has the form\\ 
\begin{equation}
G^s_{ab}\left(\bar {\nu}_L^{~a}, ~ (\bar N_R^{~c})^a\right)
\left( \begin{array}{ccc}
\omega_1 & ~ \omega_2 /\sqrt{2}\\[3mm]
\omega_2 /\sqrt{2}& ~ \omega_3
\end{array} \right)
\left(\begin{array}{c} (\nu_L^{~c})^b\\[2mm] N_R^{~b}
\end{array}\right).   
\end{equation}
\\
The latter is nothing but the familiar Dirac--Majorana mass term \\
\begin{equation}
 {1\over 2} \left(\bar {\nu}_L, ~ \bar N_R^{~c}\right)
\left( \begin{array}{ccc}
{\bf m}_T & ~ {\bf m}_D\\[3mm]
{\bf m}_D & ~ {\bf m}_S
\end{array} \right)
\left(\begin{array}{c} \nu_L^{~c}\\[2mm] N_R
\end{array}\right),    
\end{equation}
\\ 
where the family indeces are omitted and ~ ${\bf m}_{T,D,S}$ ~ are $3\times 3$ 
matrices with the following elements 
\begin{eqnarray}
({\bf m}_T)_{ab} &=& 2G^s_{ab}~\omega_1,\\[2mm]
({\bf m}_D)_{ab} &=& 2G^s_{ab}~\omega_2/\sqrt{2},\\[2mm]
({\bf m}_S)_{ab} &=& 2G^s_{ab}~\omega_3.
\end{eqnarray} 
\\
An analysis of a mass term of this kind is well known. As  the VEV's $\omega_i$ 
do not effect the charged leptons we may be able to  make some of them arbitrarily 
small to get tiny masses for neutrinos. At the seesaw limit (if allowed)
\begin{equation}
{\bf m}_T\approx 0,~~  {\bf m}_S \gg {\bf m}_D 
\end{equation}
or equivalently, 
\begin{equation}
\omega_1\approx 0,~~  \omega_3 \gg \omega_2, 
\label{seesaw}
\end{equation}
\\
we get two eigen mass matrices (family--mixing, in general)   
\begin{equation}
{\bf m}_1 = {\bf m}_D^2/{\bf m}_S, ~~ {\bf m}_2 = {\bf m}_S.
\end{equation}
\\
The condition (\ref{seesaw}) could be accepted in some circumstance, for example, 
when $\omega_3$ characterizes the energy scale of bearking $SU(3)_L$ down to
$SU(2)_L$ and therefore it must be much bigger than $\omega_2$ and $\omega_1$
characterizing the scales of breaking $SU(3)_L$ and $SU(2)_L$ down to $U(1)$: 
\begin{equation}
\omega_3 \gg \omega_2 \gg \omega_1.
\end{equation}

  Diagonalizing the matrix $G^s_{ab}$ and keepping (\ref{seesaw}) valid 
we get for each eigenstate $\alpha=1,2,3$,  a diagonalized mass matrix 
\begin{eqnarray}
(m_T)_\alpha &=& 2G^s_\alpha~\omega_1,\\[2mm]
(m_D)_\alpha &=& 2G^s_\alpha~\omega_2/\sqrt{2},\\[2mm]
(m_S)_\alpha &=& 2G^s_\alpha~\omega_3,
\label{mmdiago}
\end{eqnarray} 
leading to the Majorana neutrinos 
\begin{eqnarray}
(n_{1L})_\alpha&=& (\nu_L)_\alpha - {(m_D)_\alpha \over
(m_S)_\alpha}(\nu_R^c)_\alpha,\\[2mm]  
(n_{2L})_\alpha &=&  {(m_D)_\alpha\over (m_S)_\alpha}(\nu_L)_\alpha 
+ (\nu_R^c)_\alpha 
\end{eqnarray} 
with masses 
\begin{eqnarray}
(m_1)_\alpha &=& (m_D^2)_\alpha/(m_S)_\alpha \equiv G^s_\alpha 
{(\omega_2)^2 \over \omega_3},\\[2mm] 
(m_2)_\alpha &=& (m_S)_\alpha \equiv 2G^s_\alpha~\omega_3,  
\label{majomass}
\end{eqnarray}
where $(m)_\alpha$ is a diagonal element of a matrix ${\bf
m}$. \\

The above analysis is rough. In order to have a precise conclusion, we have to analyse the model in more details. Also, following the approaches of  \cite{akl} we can investigate the Higgs sector alone and its mass spectrum.\\[7mm]
{\bf Acknowledgements}: I would like to thank Hoang Ngoc Long, S. Petcov and A. Smirnov for discussions,  and S. Randjbar-Daemi for kind hospitality at the Abdus Salam  ICTP, Trieste, Italy.  

\end{document}